\documentclass[aps,preprint,nofootinbib,floatfix,superscriptaddress]{revtex4}
\usepackage{epsfig}
\usepackage{graphicx}
\usepackage{multirow}
\usepackage{latexsym,amsbsy,amsmath}
\usepackage{stackrel}
\def\llcol#1#2{\tilde{\lambda}_{#1}.\tilde{\lambda}_{#2}}
\begin{document}
\title{Constituent quark-model hidden-flavor pentaquarks}
\author{H.~Garcilazo} 
\email{hgarcilazos@ipn.mx} 
\affiliation{Escuela Superior de F\' \i sica y Matem\'aticas, \\ 
Instituto Polit\'ecnico Nacional, Edificio 9, 
07738 Mexico D.F., Mexico} 
\author{A.~Valcarce} 
\email{valcarce@usal.es} 
\affiliation{Departamento de F\'\i sica Fundamental,\\ 
Universidad de Salamanca, E-37008 Salamanca, Spain}
\date{\today} 

\begin{abstract}
We study hidden-flavor pentaquarks, $Q\bar Q qqq$,
based on a constituent quark-model with a standard quark-quark 
interaction that reproduces the low-energy meson and baryon spectra. 
We make use of dynamical correlations
between the heavy quarks arising from the Coulomb-like nature
of the short-range interaction. A detailed 
comparison is made with other results in the literature
and with experimental data. Our results show 
a different pattern for open-flavor and 
hidden-flavor pentaquarks, as suggested by the data.
Further implications about the existence of quarkonia bound 
to nuclei are discussed.
\end{abstract}
\maketitle

\section{Introduction}
\label{secI}
The last two decades have witnessed a significant increase in the number
of new experimental states discovered in the heavy-hadron spectra.
A number of reviews in the recent 
literature~\cite{Che16,Bri16,Ric16,Hos16,Che17,Leb17,Ali17,Esp17,Guo18,Ols18,Kar18,Bra20} 
have summarized both experimental and theoretical developments.
As a general conclusion it has emerged the idea 
the heavy-hadron spectra shows the
contribution of states that do not belong to the simplest quark-antiquark 
(meson) or three-quark (baryon) structures proposed by Gell-Mann~\cite{Gel64}. 
This is quite evident by the recent discovery of double heavy tetraquarks 
with manifestly exotic quantum numbers~\cite{Aai21,Aaj21}. However, most of the 
intriguing experimental states have ordinary quantum numbers, which 
suggests that they could correspond to more sophisticated quark structures 
allowed by QCD.

The new experimental findings have given rise to a substantial theoretical effort 
to understand the spectroscopy and structure of these novel states.
Different proposals have been studied with their benefits but also 
drawbacks: hadronic molecules, diquarks, hadroquarkonium, hybrids, 
kinematical threshold effects,... -- see the reviews above 
for a detailed summary. No single theoretical model has 
emerged to give the big picture. A full understanding might 
require incorporating several relevant possibilities, perhaps with a
different mix for every state.

A major question lying behind the emerging pattern in the heavy-hadron 
spectra is whether or not hadrons with a more sophisticated quark substructure, the 
so-called multiquarks, could be observed in nature.\footnote{In the case of 
hadrons with manifestly
exotic quantum numbers the recent experimental discoveries~\cite{Aai21,Aaj21}
deliver a positive answer to this question. For hadrons with non-exotic quantum numbers
this is a long-standing open-ended question~\cite{Jaf77,Jae77}.} Multiquarks, 
considered either as compact states or hadronic molecules, have been 
the focus of much writing over the past two decades.\footnote{Broadly speaking,
in a constituent quark language, hadronic molecules are a particular 
case of multiquarks, those that are composed of a certain number 
of conventional hadrons~\cite{Gel64}.} In atomic or nuclear physics 
the development of bound states relies on the existence of attractive 
enough interactions in channels without tight constraints imposed by 
the Pauli principle. In contrast, dealing with the quark substructure
the color degree of freedom comes to play a key role to yield bound 
states. Multiquarks (tetraquarks, pentaquarks and so on) do always 
contain substructures made of color singlets but, in contrast to
atomic and nuclear physics, they could also be dominantly 
made of structures that are not allowed to exist isolated in nature.

The simplest quark structures proposed by Gell-Mann~\cite{Gel64}
could only decay strongly through the breaking of the color flux 
tube generating other color singlet hadrons. Regarding non-ordinary 
hadrons (multiquarks) with standard quantum numbers the most salient 
feature is the scarcity of bound states, restricted to very
peculiar configurations. This is concluded both in lattice QCD 
approaches~\cite{Hug18,Hud20} and in constituent models~\cite{Sil93,Ric18}
provided that there are no restrictions other than those imposed by
the Pauli principle.
The difficulty to encounter multiquark hadrons that do not immediately break 
into their fall-apart decay has suggested the use of
correlations among the constituents due to a 
more complex dynamics that, for instance, might restrict the
quantum numbers of the internal substructures.   
In this line of thoughts have emerged, among others, the so-called 
diquark models~\cite{Ans93,Fre82}, where the color degree of freedom 
of two quarks (antiquarks) is frozen to a particular state.\footnote{There
are different alternatives for diquark structures in the literature, but
they all share constraints in the color quantum numbers of 
pairs of the constituents. For instance, there are studies where the 
color of a couple of quarks is restricted to a 
${\bf \bar 3}$ state~\cite{Mai15,Gir19,Ali19,Shi21} or others where the color
of a quark-antiquark pair is taken to be only a ${\bf 8}$ state~\cite{Wul17}.}
In contrast to uncorrelated multiquark models, a larger number of 
theoretical non-ordinary hadrons appears. We refer the reader to 
Refs.~\cite{Ans93,Fre82} for advantages and/or disadvantages of 
the so-called diquark approximation.

In this paper we explore a theoretical 
scenario where the dynamics of a multiquark system
remains marked by correlations between
heavy flavors dictated by QCD~\cite{Jaf05}. 
For this reason we have chosen hidden-flavor  
pentaquarks for our study, i.e., $Q\bar Q qqq$. 
The theoretical pattern obtained will be an additional tool
for analyzing the growing number 
of states in the quarkonium-nucleon 
energy region. As discussed below, the correlations
between the heavy flavors turn the five-body
problem into a more tractable three-body problem. 
Our study is based on a constituent quark model that 
has often been used for exploratory studies,
whose results have been refined and confirmed 
by more rigorous treatments of QCD. 
For instance, the recently discovered 
flavor-exotic mesons, $T^+_{cc}\equiv cc\bar u\bar d$~\cite{Aai21,Aaj21}, were first 
predicted by potential-model calculations~\cite{Ade82} and later reinforced 
by more refined potential-model calculations, lattice simulations and QCD sum 
rules~\cite{Fra17,Kar17,Eic17,Bic16,Jun19,Pad22,Luo17,Duc13,Car11,Cza18,Jan04}.

The structure of the paper is the following. In the next section
we present the model. We will show the interacting
potential between quarks and the Hilbert
space arising from the correlations between the heavy flavors.
Sec.~\ref{secIII} is devoted to discuss the solution of 
the Faddeev equations for the bound state three-body 
problem considering the coupling among all two-body 
amplitudes. In Sec.~\ref{secIV} we present and discuss our 
results compared to those of other constituent model
studies and experimental data. Finally, our 
conclusions are summarized in Sec.~\ref{secV}.

\section{Dynamical model}
\label{secII}
We study the hidden-flavor pentaquarks, $Q\bar Q qqq$, arising
from dynamical correlations between the heavy flavors.
Much has been learned about the outcome of the so-called diquark 
picture~\cite{Mai15,Gir19,Ali19,Shi21}. In the case of tetraquarks, it means to
model the system as a bound color-${\bf \bar 3}$ diquark and a bound color-${\bf 3}$ 
antidiquark. In other words, the color ${\bf 6} {\bf \bar 6}$ component
is not considered. Possible pentaquarks with configurations where the $Q \bar Q$ pair 
is a color octet have also been explored~\cite{Wul17}.
Needless to say, if a multiquark contains color configurations that 
are not present asymptotically in the thresholds, this could be a basic ingredient 
which may lead to bound states. 

\begin{figure}[t]
\vspace*{-0.7cm}
\includegraphics[width=0.75\columnwidth]{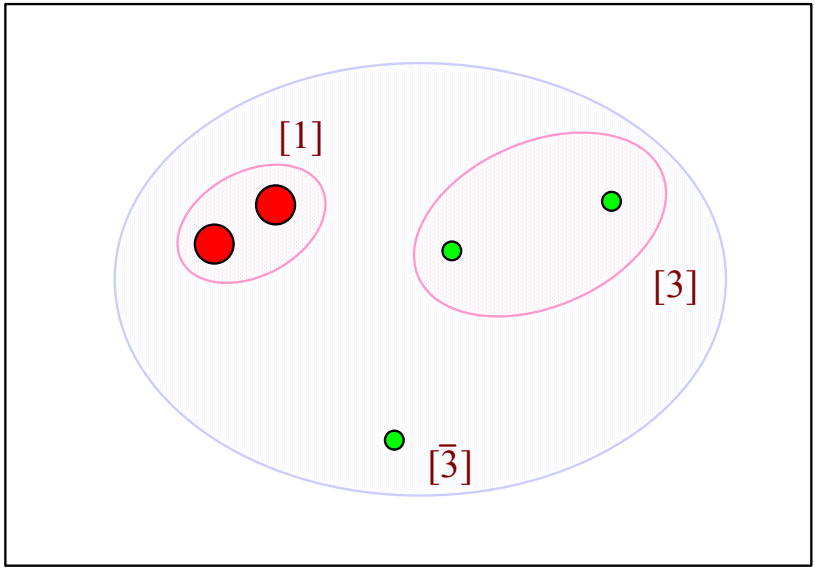}
\vspace*{-9.cm}
\caption{Color structure of a hidden-flavor pentaquark arising from the color Coulomb-like
correlations between the heavy flavors. The large red circles stand for the heavy 
quark-antiquark pair, the small green circles represent the light quarks. The 
numbers between squared brackets denote the color wave function.}
\label{fig1}
\end{figure}

The idea behind these approaches is to select the most favorable configurations
to generate stable multiquarks. For example, the diquark models of 
Refs.~\cite{Mai15,Gir19,Ali19,Shi21} are based on the fact
that a color-${\bf \bar 3}$ $qq$ state is an attractive channel 
whereas the color-${\bf 6}$ is repulsive. In the same vein, a 
color-${\bf 1}$ $q\bar q$ state is an attractive channel
whereas the color-${\bf 8}$ is repulsive. Working at 
leading order with a $Q\bar Qqqq$ pentaquark, neglecting the spin-spin interaction, 
if a $Qq$ color-${\bf \bar 3}$ diquark has a binding proportional to
$m_q$, in the same units the $Q\bar Q$ color-${\bf 1}$ system 
has a binding proportional to $2M_Q$. Therefore, the color Coulomb-like interaction 
between the components of a hidden-flavor pentaquark 
favors a $Q\bar Q$ color singlet instead of a color octet, as emphasized in Ref.~\cite{Jaf05}. 
As a consequence, the color wave function of a pentaquark 
would be uniquely determined, see Fig.~\ref{fig1}, and would be given by,
\begin{equation}
\Psi^{\rm Color}_{\rm Pentaquark} \,\, = \,\, {\bf 3}_q \, \otimes \, {\bf 1}_{(Q \bar Q)} \, \otimes \, {\bf \bar 3}_{(qq)} \, ,
\label{ecu1}
\end{equation}
thus avoiding the repulsive component originating from the color octet of the
heavy quark-antiquark pair. It is worth noting that the constraints imposed
by the color Coulomb-like nature of the short-range interaction between the heavy flavors 
arise naturally in constituent quark-model based studies of double 
heavy tetraquarks~\cite{Her20,Men21}.

Taking into account that the heavy quarks have isospin zero and the antisymmetric character
of the color-${\bf \bar 3}$ $qq$ wave function -- what implies that its spin
and isospin must be identical -- one can identify the different vectors that contribute to
any $(I,J)$ pentaquark for the lowest lying states, i.e., in the case of a fully 
symmetric radial wave function,
\begin{equation}
\Psi^{(I,J)}_{\rm Pentaquark} \,\, = \,\, \{{\bf 3}_c, i_1=1/2, s_1=1/2 \}_{q} \, \otimes \, 
\{{\bf 1}_c, i_2=0, s_2 \}_{(Q \bar Q)} \, \otimes \, 
\{{\bf \bar 3}_c, i_3=s_3, s_3\}_{(qq)} \, .
\label{ecu2}
\end{equation}
We summarize in Table~\ref{tab1} the possible value of the quantum
numbers leading to an allowed $(I,J)$ hidden-flavor pentaquark. $s_1$ stands for the spin
of the single light quark (with isospin $1/2$), $s_2$ denotes the spin of the
heavy quark-antiquark pair (with isospin zero) and finally $s_3$ represents the spin of the
light quark pair (with the restrictions imposed by the Pauli principle such that $s_3=i_3$).
The notation in the last column will be used in the next sections to identify the wave 
function of the different pentaquarks.
\begin{table}[t]
\begin{tabular}{cp{0.5cm}cp{1cm}cp{0.35cm}cp{0.35cm}cp{0.35cm}cp{0.35cm}c} \hline\hline
         $I$             && $J$     && $s_1$   && $s_2$    && $s_3$  && Vector \\ \hline
\multirow{7}{*}{$1/2$}   && $1/2$   && \multirow{7}{*}{$1/2$}   && $0$      && $0$    &&  $v_1$ \\
                         && $1/2$   &&    && $1$      && $0$    &&  $v_2$ \\
                         && $1/2$   &&    && $0$      && $1$    &&  $v_3$ \\
                         && $1/2$   &&    && $1$      && $1$    &&  $v_4$ \\
                         && $3/2$   &&    && $1$      && $0$    &&  $w_1$ \\
                         && $3/2$   &&    && $1$      && $1$    &&  $w_2$ \\ 
                         && $5/2$   &&    && $1$      && $1$    &&  $y_1$ \\ 												
												\hline
\multirow{4}{*}{$3/2$}   && $1/2$   && \multirow{4}{*}{$1/2$}   && $1$      && $1$    &&  $v_4$ \\ 
                         && $3/2$   &&    && $0$      && $1$    &&  $w_3$ \\
                         && $3/2$   &&    && $1$      && $1$    &&  $w_2$ \\                   
                         && $5/2$   &&    && $1$      && $1$    &&  $y_1$ \\ \hline \hline
\end{tabular}
\caption{Quantum numbers of the different channels contributing to a given
$(I,J)$ hidden-flavor pentaquark according to Eq.~(\ref{ecu2}). See text for details.} 
\label{tab1}
\end{table}

Once the Hilbert space arising from the correlation between the heavy flavors has been
delimited, the only ingredient left for our study is a realistic interaction between the quarks.
In this paper we adopt a generic constituent model, containing chromoelectric and 
chromomagnetic contributions, tuned to reproduce the masses of the mesons and baryons 
entering the various vectors shown in Table~\ref{tab1}.
We adopt the so-called AL1 model by Semay and Silvestre-Brac~\cite{Sem94}, widely used in 
a number of exploratory studies of multiquark systems~\cite{Sil93,Ric18,Jan04,Her20,Ric17,Hiy18,Meg19}. It includes a standard 
Coulomb-plus-linear central potential, supplemented by a smeared version of the chromomagnetic interaction,
\begin{align}
\label{ecu3}
V(r)  & =  -\frac{3}{16}\, \llcol{i}{j}
\left[\lambda\, r - \frac{\kappa}{r}-\Lambda + \frac{V_{SS}(r)}{m_i \, m_j}  \, \vec \sigma_i \cdot \vec\sigma_j\right] \, ,\\ \nonumber \\
V_{SS}  &= \frac{2 \, \pi\, \kappa^\prime}{3\,\pi^{3/2}\, r_0^3} \,\exp\left(- \frac{r^2}{r_0^2}\right) ~,\quad
 r_0 =  A \left(\frac{2 m_i m_j}{m_i+m_j}\right)^{-B} \, , \nonumber
\end{align}
where
$\lambda=$ 0.1653 GeV$^2$, $\Lambda=$ 0.8321 GeV, $\kappa=$ 0.5069, $\kappa^\prime=$ 1.8609,
$A=$ 1.6553 GeV$^{B-1}$, $B=$ 0.2204, $m_u=m_d=$ 0.315 GeV, $m_s=$ 0.577 GeV, $m_c=$ 1.836 GeV and $m_b=$ 5.227 GeV. 
Here, $\llcol{i}{j}$ is a color factor, suitably modified for the quark-antiquark pairs.
Note that the smearing parameter of the spin-spin 
term is adapted to the masses involved in the quark-quark or quark-antiquark pairs. 
The parameters of the AL1 potential are constrained in a 
simultaneous fit of 36 well-established mesons and 53 baryons,
with a remarkable agreement with data, as could be seen in Table~2 of Ref.~\cite{Sem94}.
It is worth to note that although the $\chi^2$ obtained in Ref.~\cite{Sem94} with the AL1 potential
is slightly larger than the one obtained with other models, this is essentially
because a number of resonances with high angular momenta were considered. The AL1 model
is very well suited to study the low-energy hadron spectra~\cite{Sil96}. 
The spin-color algebra of the five-quark system has been worked elsewhere~\cite{Ale11,Ric17}.
The capability of the model to describe relevant ordinary hadrons: $Q\bar Q$, $q\bar Q$, $qqq$ and $Qqq$,
is illustrated in Table~\ref{tab2} for $Q=c$.
\begin{table}[t]
\begin{tabular}{cp{0.3cm}cp{0.1cm}cp{1cm}cp{0.3cm}cp{0.1cm}c}
\hline\hline
\multicolumn{5}{c}{Baryons} && \multicolumn{5}{c}{Mesons} \\ \hline
State                   && AL1     &&  Exp.   && State                   &&  AL1     &&  Exp.  \\ \hline
$N$		                  &&   996  &&   940  && $D$                     &&  1862   &&  1868  \\
$\Delta$                &&  1307  &&  1232  && $D^*$                   &&  2016   &&  2008  \\
$\Lambda_c$             &&  2292  &&  2286  && $\eta_c$	               &&  3005   &&  2989  \\
$\Sigma_c$              &&  2467  &&  2455  && $J/\psi$                &&  3101   &&  3097  \\
$\Sigma^*_c$            &&  2546  &&  2518  &&                         &&         &&         \\
\hline \hline
\end{tabular}
\caption{Masses (in MeV) of ordinary hadrons 
calculated with the AL1 potential of Eq.~(\ref{ecu3}),
compared to the experimental values.}
\label{tab2}
\end{table}

The bound nature of a multiquark could arise from an attractive medium-long range
interaction generated by the exchange of color-singlet Goldstone 
bosons~\cite{Vij04,Hua16,Yan17}.\footnote{A detailed discussion of the relative role
played by the one-gluon exchange with respect to the Goldstone boson exchanges in a hybrid
constituent quark model to lead to stable tetraquarks can be found in Ref.~\cite{Vij04}.}
In addition, it is known that the short-range one-gluon exchange interaction
generates a strong repulsive force in the $NN$ $S$-wave partial waves.
This feature is not universal for a hadron-hadron interaction in general.
Indeed, it disappears for some channels of, among others, the $\Delta\Delta$ and $N\Delta$ systems,
generating a positive phase shift at low energies~\cite{Oka80}.\footnote{A positive phase shift is
an indication of an attractive interaction. If it goes above $\pi/2$ degrees and returns to zero
is a sign of a resonance and if it goes to $\pi$ degrees at zero energy it 
shows the existence of a bound state.}
This attractive short-range behavior was the basis of resonances predicted in the 
$\Delta\Delta$ and $N\Delta$ systems~\cite{Gol89,Pan01,Mot99,Val01}, some of which have been 
established experimentally~\cite{Cle17}. Thus, the short-range quark-gluon dynamics 
of multiquark systems may also induce stability. 
		
Let us finally note that once the color wave function is frozen 
by the dynamical correlations between the heavy flavors and being
all the constituents spin $1/2$ particles, the flavor-independence 
of the interacting potential makes the five-body problem to factorize 
into the three-body problem shown in Fig.~\ref{fig1}. The existence of 
correlated substructures in a many-quark system 
leads, in general, to more tractable technical problems.
This is for instance the case of tetraquark studies under the
diquark hypothesis~\cite{Kar17,Eic17}, where the four-body problem is reduced to a two-body
problem of effective diquarks with a mass fixed in other known hadron sectors. 
In the problem addressed in this manuscript, the correlations between the heavy
flavors leads to a three-body problem that can be exactly solved
by means of the Faddeev equations. This also allows to overcome the 
difficulties associated to the minimization procedure inherent
to variational methods for getting fully converged results.
This is particularly relevant working close to 
open thresholds. We discuss in the next section the 
solution of the Faddeev equations for the bound 
state problem of a three-body system. 

\section{The three-body problem}
\label{secIII}
The freezing of the color wave function described in the previous section 
leads to the effective three-body problem shown in Fig.~\ref{fig1}
and summarized in the wave function of Eq.~(\ref{ecu2}). The allowed spin and isospin
values of the different particles -- $1$: light quark, $2$: heavy quark-antiquark pair, $3$:
two light-quark pair -- are indicated in Table~\ref{tab1}.
\begin{table}[t]
\caption{Channels that are coupled together for the different $J=1/2$
states, $v_i$ in Table~\ref{tab1}. See text for details.}
\begin{tabular}{cp{0.5cm}cp{0.3cm}cp{0.3cm}cp{0.3cm}cp{0.3cm}cp{0.5cm}c}
 \hline\hline
                      && $s_1$ && $s_2$ && $S_3$    && $s_3$&& $ I $     && $ F $           \\ \hline
$v_1$                 && $1/2$ && $ 0 $ && $ 1/2  $ && $ 0 $&& $1/2$     && $9/8$           \\
$v_2$                 && $1/2$ && $ 1 $ && $ 1/2  $ && $ 0 $&& $1/2$     && $3/8$           \\
$v_3$                 && $1/2$ && $ 0 $ && $ 1/2  $ && $ 1 $&& $1/2$     && $9/8$           \\
\multirow{2}{*}{$v_4$}&& $1/2$ && $ 1 $ && $ 1/2  $ && $ 1 $&& $1/2,3/2$ && $3/8$           \\
                      && $1/2$ && $ 1 $ && $ 3/2  $ && $ 1 $&& $1/2,3/2$ && $3/2\sqrt{2}$   \\
\cline{1-9}
                      && $s_2$ && $s_3$ && $S_1$    && $s_1$&&           &&           \\ \cline{1-9}
$v_1$                 && $ 0 $ && $ 0 $ && $  0   $ && $1/2$&& $1/2$     && $ 0 $      \\
$v_2$                 && $ 1 $ && $ 0 $ && $  1   $ && $1/2$&& $1/2$     && $ 0 $      \\
$v_3$                 && $ 0 $ && $ 1 $ && $  1   $ && $1/2$&& $1/2$     && $ 0 $      \\
\multirow{2}{*}{$v_4$}&& $ 1 $ && $ 1 $ && $  0   $ && $1/2$&& $1/2,3/2$ && $ -4/3 $   \\
                      && $ 1 $ && $ 1 $ && $  1   $ && $1/2$&& $1/2,3/2$ && $ -5/3 $   \\
\cline{1-9}
                      && $s_3$ && $s_1$ && $S_2$    && $s_2$&&           &&                \\ \cline{1-9}
$v_1$                 && $ 0 $ && $1/2$ && $ 1/2  $ && $ 0 $&& $1/2$     && $9/8$           \\
$v_2$                 && $ 0 $ && $1/2$ && $ 1/2  $ && $ 1 $&& $1/2$     && $9/8$           \\
$v_3$                 && $ 1 $ && $1/2$ && $ 1/2  $ && $ 0 $&& $1/2$     && $3/8$           \\
\multirow{2}{*}{$v_4$}&& $ 1 $ && $1/2$ && $ 1/2  $ && $ 1 $&& $1/2,3/2$ && $3/8$           \\
                      && $ 1 $ && $1/2$ && $ 3/2  $ && $ 1 $&& $1/2,3/2$ && $3/2\sqrt{2}$   \\
\hline\hline
\end{tabular}
\label{tab4} 
\end{table}
\begin{table}[t]
\caption{Same as Table~\ref{tab4} for $J=3/2$ states, $w_i$ in Table~\ref{tab1}.}
\begin{tabular}{cp{0.5cm}cp{0.3cm}cp{0.3cm}cp{0.3cm}cp{0.3cm}cp{0.5cm}c}
 \hline\hline
                      && $s_1$ && $s_2$ && $S_3$    && $s_3$&& $ I $     && $ F $           \\ \hline
$w_1$                 && $1/2$ && $ 1 $ && $ 3/2  $ && $ 0 $&& $1/2$     && $3/2\sqrt{2}$    \\
\multirow{2}{*}{$w_2$}&& $1/2$ && $ 1 $ && $ 1/2  $ && $ 1 $&& $1/2,3/2$ && $3/8$           \\
                      && $1/2$ && $ 1 $ && $ 3/2  $ && $ 1 $&& $1/2,3/2$ && $3/2\sqrt{2}$   \\
$w_3$                 && $1/2$ && $ 0 $ && $ 1/2  $ && $ 1 $&& $3/2$     && $9/8$           \\											
\cline{1-9}
                      && $s_2$ && $s_3$ && $S_1$    && $s_1$&&           &&            \\ \cline{1-9}
$w_1$                 && $ 1 $ && $ 0 $ && $  1   $ && $1/2$&& $1/2$     && $ 0 $      \\

\multirow{2}{*}{$w_2$}&& $ 1 $ && $ 1 $ && $  1   $ && $1/2$&& $1/2,3/2$ && $ 2/3 $   \\
                      && $ 1 $ && $ 1 $ && $  2   $ && $1/2$&& $1/2,3/2$ && $ -2/3 $   \\
$w_3$                 && $ 0 $ && $ 1 $ && $  1   $ && $1/2$&& $3/2$     && $ 0 $      \\										
\cline{1-9}
                      && $s_3$ && $s_1$ && $S_2$    && $s_2$&&           &&                \\ \cline{1-9}
$w_1$                 && $ 0 $ && $1/2$ && $ 1/2  $ && $ 1 $&& $1/2$     && $9/8$           \\
\multirow{2}{*}{$w_2$}&& $ 1 $ && $1/2$ && $ 1/2  $ && $ 1 $&& $1/2,3/2$ && $3/8$           \\
                      && $ 1 $ && $1/2$ && $ 3/2  $ && $ 1 $&& $1/2,3/2$ && $3/2\sqrt{2}$   \\
$w_3$                 && $ 1 $ && $1/2$ && $ 3/2  $ && $ 0 $&& $3/2$     && $3/2\sqrt{2}$   \\											
\hline\hline
\end{tabular}
\label{tab5} 
\end{table}
\begin{table}[t]
\caption{Same as Table~\ref{tab4} for $J=5/2$ states, $y_i$ in Table~\ref{tab1}.}
\begin{tabular}{cp{0.5cm}cp{0.3cm}cp{0.3cm}cp{0.3cm}cp{0.3cm}cp{0.5cm}c}
 \hline\hline
                      && $s_1$ && $s_2$ && $S_3$    && $s_3$&& $ I $     && $ F $           \\ \hline
$y_1$                 && $1/2$ && $ 1 $ && $ 3/2  $ && $ 1 $&& $1/2,3/2$ && $3/2\sqrt{2}$    \\
\cline{1-9}
                      && $s_2$ && $s_3$ && $S_1$    && $s_1$&&           &&           \\ \cline{1-9}
$y_1$                 && $ 1 $ && $ 1 $ && $  2   $ && $1/2$&& $1/2,3/2$ && $ 1 $      \\
\cline{1-9}
                      && $s_3$ && $s_1$ && $S_2$    && $s_2$&&           &&                 \\ \cline{1-9}
$y_1$                 && $ 1 $ && $1/2$ && $ 3/2  $ && $ 1 $&& $1/2,3/2$ && $3/2\sqrt{2}$    \\
\hline\hline
\end{tabular}
\label{tab6} 
\end{table}

Three-body states in which a particle has a given spin can only
couple to other three-body states in which that particle has the same spin, 
since the spinors corresponding to different eigenvalues are orthogonal.
This is shown in detail in Appendix~\ref{AppA}.
The same applies for isospin. This leads to a decoupling of the integral 
equations in various sets in which the spin and isospin of each particle 
remains the same. We show the different sets contributing to
$J=1/2$, $3/2$, and $5/2$ in Tables~\ref{tab4},~\ref{tab5}, 
and~\ref{tab6}, respectively. Besides the notation introduced in Table~\ref{tab1},
we denote by $S_i$ and $I_i$ the spin and isospin of the pair $jk$. As discussed 
above, the isospin of each particle is determined once the spin is given, so
it is not shown in the tables. Finally, $F$ is the expectation value of
the $\vec \sigma_i \cdot \vec\sigma_j$ operator, responsible for
the coupling of different two-body amplitudes as explained below.

To solve the Faddeev equations in momentum space for the case of
confining potentials we follow the method developed in Ref.~\cite{Gar03},
that it is described below for $S$- and $P$-wave states.

\subsection{$S$-wave states}
\label{subsecIII.1}
We restrict ourselves to the configurations where all three 
particles are in $S$-wave states so that the Faddeev equations 
for the bound-state problem with total
isospin $I$ and total spin $J$ are,
\begin{eqnarray}
T_{i;IJ}^{I_iS_i}(p_iq_i) = &&\sum_{j\ne i}\sum_{I_jS_j}
\frac{1}{2}\int_0^\infty q_j^2dq_j
\int_{-1}^1d{\rm cos}\theta\, 
t_{i;I_iS_i}(p_i,p_i^\prime;E-q_i^2/2\nu_i) 
\nonumber \\ &&
\times \,
h_{ij;IJ}^{I_iS_i;I_jS_j}
\frac{1}{E-p_j^2/2\eta_j-q_j^2/2\nu_j}\;
T_{j;IJ}^{I_jS_j}(p_jq_j) \, , 
\label{eq1} 
\end{eqnarray}
where $t_{i;I_iS_i}$ stands for the two-body amplitudes
with isospin $I_i$ and spin $S_i$ and
$\eta_i$ and $\nu_i$ are the corresponding reduced masses,
\begin{eqnarray}
\eta_i &=& \frac{m_jm_k}{m_j+m_k} \, , \nonumber\\
\nu_i &=& \frac{m_i(m_j+m_k)}{m_i+m_j+m_k} \, ,
\label{eq3}
\end{eqnarray}
$\vec p_i^{\; \prime}$
is the momentum of the pair $jk$ (with $ijk$ an even permutation of
$123$) and $\vec p_j$ is the momentum of the pair
$ki$  which are given by,
\begin{eqnarray}
\vec p_i^{\; \prime} &=& -\vec q_j-\alpha_{ij}\vec q_i \, , \nonumber\\
\vec p_j &=& \vec q_i+\alpha_{ji}\vec q_j \, ,
\label{eq3p}
\end{eqnarray}
where,
\begin{eqnarray}
\alpha_{ij} &=& \frac{\eta_i}{m_k} 
\, , \nonumber\\
\alpha_{ji} &=& \frac{\eta_j}{m_k} 
\, ,
\label{eq3pp}
\end{eqnarray}
so that,
\begin{eqnarray}
p_i^\prime &=& \sqrt{q_j^2+\alpha_{ij}^2q_i^2+2\alpha_{ij}
q_iq_j{\rm cos}\theta} \, , \nonumber \\
p_j &=& \sqrt{q_i^2+\alpha_{ji}^2q_j^2+2\alpha_{ji}
q_iq_j{\rm cos}\theta} \, .
\label{eq5}
\end{eqnarray}
$h_{ij;IJ}^{I_iS_i;I_jS_j}$ are the spin--isospin coefficients,
\begin{eqnarray}
h_{ij;IJ}^{I_iS_i;I_jS_j}= &&
(-)^{I_j+i_j-I}\sqrt{(2I_i+1)(2I_j+1)}
W(i_ji_kIi_i;I_iI_j)
\nonumber \\ && \times
(-)^{S_j+s_j-J}\sqrt{(2S_i+1)(2S_j+1)}
W(s_js_kJs_i;S_iS_j) \, , 
\label{eq6}
\end{eqnarray}
where $W$ is a Racah coefficient and $i_i$, $I_i$, and $I$ 
($s_i$, $S_i$, and $J$) are the isospins (spins) of particle $i$,
of the pair $jk$, and of the three--body system.

In Eq.~(\ref{eq1}) the variable $p_i$ runs from 0 to $\infty$. Thus, 
it is convenient to make the transformation,
\begin{equation}
x_i=\frac{p_i-b}{p_i+b} \, ,
\label{eq7}
\end{equation}
where the new variable $x_i$ runs from $-1$ to $1$, and $b$ is a scale
parameter that has no effect on the solution. With this transformation
Eq.~(\ref{eq1}) takes the form,
\begin{eqnarray}
T_{i;IJ}^{I_iS_i}(x_iq_i) = &&\sum_{j\ne i}\sum_{I_jS_j}
\frac{1}{2}\int_0^\infty q_j^2dq_j
 \int_{-1}^1d{\rm cos}\theta\; 
t_{i;I_iS_i}(x_i,x_i^\prime;E-q_i^2/2\nu_i) 
\nonumber \\ &&
\times \,
h_{ij;IJ}^{I_iS_i;I_jS_j}
\frac{1}{E-p_j^2/2\eta_j-q_j^2/2\nu_j}\;
T_{j;IJ}^{I_jS_j}(x_jq_j) \, . 
\label{eq8} 
\end{eqnarray}
Since the variables $x_i$ and $x_i^\prime$ run from $-1$ to $1$, one can 
expand the amplitude $t_{i;I_iS_i}(x_i,x_i^\prime;e)$
in terms of Legendre polynomials as,
\begin{equation}
t_{i;I_iS_i}(x_i,x_i^\prime;e)=\sum_{nr}P_n(x_i)\tau_{i;I_iS_i}^{nr}(e)P_r(x'_i) \, ,
\label{eq9}
\end{equation}
where the expansion coefficients are given by,
\begin{equation}
\tau_{i;I_iS_i}^{nr}(e)= \frac{2n+1}{2}\frac{2r+1}{2}\int_{-1}^1dx_i
\int_{-1}^1 dx_i^\prime\; P_n(x_i) 
t_{i;I_iS_i}(x_i,x_i^\prime;e)P_r(x_i^\prime) \, .
\label{eq10} 
\end{equation}
Applying expansion~(\ref{eq9}) in Eq.~(\ref{eq8}) one gets,
\begin{equation}
T_{i;IJ}^{I_iS_i}(x_iq_i) = \sum_n P_n(x_i) T_{i;IJ}^{nI_iS_i}(q_i) \, ,
\label{eq11}
\end{equation}
where $T_{i;IJ}^{nI_iS_i}(q_i)$ satisfies the one-dimensional integral equation,
\begin{equation}
T_{i;IJ}^{nI_iS_i}(q_i) = \sum_{j\ne i}\sum_{mI_jS_j}
\int_0^\infty dq_j K_{ij;IJ}^{nI_iS_i;mI_jS_j}(q_i,q_j;E)\;
T_{j;IJ}^{mI_jS_j}(q_j) \, , 
\label{eq12}
\end{equation}
with
\begin{eqnarray}
K_{ij;IJ}^{nI_iS_i;mI_jS_j}(q_i,q_j;E)= &&
\sum_r\tau_{i;I_iS_i}^{nr}(E-q_i^2/2\nu_i)
\frac{q_j^2}{2}
\nonumber \\ &&
\times\int_{-1}^1 d{\rm cos}\theta\;
h_{ij;IJ}^{I_iS_i;I_jS_j}
\frac{P_r(x_i^\prime)P_m(x_j)} 
{E-p_j^2/2\eta_j-q_j^2/2\nu_j} \, .
\label{eq13} 
\end{eqnarray}
The three amplitudes $T_{1;IJ}^{rI_1S_1}(q_1)$, $T_{2;IJ}^{mI_2S_2}(q_2)$,
and $T_{3;IJ}^{nI_3S_3}(q_3)$ in Eq.~(\ref{eq12}) are coupled together.

\subsection{$P$-wave states}
\label{subsecIII.2}
In all the previous sets of coupled equations we have assumed only $S$-wave 
states. We thought interesting, however, to look into excited 
states containing one unit of orbital angular momentum. For that 
purpose we have chosen the state $v_1$ in Table~\ref{tab1}, where
$s_2=s_3=0$. We show in Table~\ref{tab7} the two-body channels that
are coupled together in this case. $\ell_i$ is the relative orbital 
angular momentum of the pair $jk$ and $\lambda_i$ is the relative orbital 
angular momentum between particle $i$ and the pair $jk$.
\begin{table}[t]
\caption{Channels that are coupled together for the
vector $v_1$ in Table~\ref{tab1} with a unit of orbital angular momentum, such that $J=1/2$
and $3/2$.}
\begin{tabular}{cp{0.5cm}cp{0.3cm}cp{0.3cm}cp{0.3cm}cp{0.3cm}cp{0.5cm}cp{0.5cm}cp{0.5cm}c}
 \hline\hline
                      && $s_1$ && $s_2$ && $S_3$    && $s_3$&& $\ell_3$ && $\lambda_3$ && $ I $     && $ F $           \\ \hline
\multirow{2}{*}{$v_1$}&& $1/2$ && $ 0 $ && $ 1/2  $ && $ 0 $&& $ 0 $    && $ 1 $       && $1/2$     && $9/8$    \\
                      && $1/2$ && $ 0 $ && $ 1/2  $ && $ 0 $&& $ 1 $    && $ 0 $       && $1/2$     && $9/8$    \\
\cline{1-13}
                      && $s_2$ && $s_3$ && $S_1$    && $s_1$&& $\ell_1$ && $\lambda_1$ && $ I $     && $ F $           \\ \cline{1-13}
\multirow{2}{*}{$v_1$}&& $ 0 $ && $ 0 $ && $ 0 $    && $1/2$&& $ 0 $    && $ 1 $       && $1/2$     && $ 0 $    \\
                      && $ 0 $ && $ 0 $ && $ 0 $    && $1/2$&& $ 1 $    && $ 0 $       && $1/2$     && $ 0 $    \\
\cline{1-13}
                      && $s_3$ && $s_1$ && $S_2$    && $s_2$&& $\ell_2$ && $\lambda_2$ && $ I $     && $ F $           \\ \cline{1-13}
\multirow{2}{*}{$v_1$}&& $ 0 $ && $1/2$ && $ 1/2  $ && $ 0 $&& $ 0 $    && $ 1 $       && $1/2$     && $9/8$    \\
                      && $ 0 $ && $1/2$ && $ 1/2  $ && $ 0 $&& $ 1 $    && $ 0 $       && $1/2$     && $9/8$    \\
\hline\hline
\end{tabular}
\label{tab7} 
\end{table}
 
To solve the integral equations~(\ref{eq1}) with one unit of orbital angular momentum
we write them symbolically as,
\begin{equation}
T_i=t_ih_{ij}G_0T_j,
\label{eq14}
\end{equation}
that has to be generalized to a matrix equation,
\begin{equation}
\begin{pmatrix}T_i^{01} \\ T_i^{10}\end{pmatrix}
=\begin{pmatrix}t_i^0 \\ t_i^1\end{pmatrix}h_{ij}G_0
\begin{pmatrix}\hat q_i\cdot\hat q_j & \hat q_i\cdot\hat p_j \\
\hat p_i^{\;\prime}\cdot\hat q_j & \hat p_i^{\;\prime}\cdot\hat p_j 
\end{pmatrix}
\begin{pmatrix}T_j^{01} \\ T_j^{10}\end{pmatrix},
\label{eq15}
\end{equation}
where, from Eq.~(\ref{eq3p}), 
\begin{eqnarray}
\hat q_i\cdot\hat q_j &=& {\rm cos}\theta \, ,
\nonumber \\
\hat q_i\cdot\hat p_j &=& \frac{q_i^2+\alpha_{ji}q_iq_j{\rm cos}\theta}{q_ip_j} \, ,
\nonumber \\
\hat p_i^{\;\prime}\cdot\hat q_j &=& \frac{-q_j^2-\alpha_{ij}q_iq_j
{\rm cos}\theta} {p_i^\prime q_j} \, ,
\nonumber \\
\hat p_i^{\;\prime}\cdot\hat p_j &=& \frac{-(1+\alpha_{ij}\alpha_{ji})q_iq_j
{\rm cos}\theta-\alpha_{ji}q_j^2-\alpha_{ij}q_i^2}{p_i^\prime p_j} \, ,
\label{eq16}
\end{eqnarray}
and $p_i^\prime$ and $p_j$ are given by Eq.~(\ref{eq5}).

\subsection{Coupling between two-body amplitudes}
\label{subsecIII.3}
In general, the two-body amplitudes that appear in Tables~\ref{tab4},~\ref{tab5},~\ref{tab6}, and~\ref{tab7} are obtained
by solving the Lippmann-Schwinger equation,
\begin{equation}
t=V+VG_0t \, ,
\label{eq17}
\end{equation}
where $V$ is the interaction given by Eq.~(\ref{ecu3}).
Due to the reduction from five to three particles, some pairs of
two-body amplitudes are coupled together. Such is the case of the
$S_1=0$ and $S_1=1$ amplitudes of the $v_4$ vector in Table~\ref{tab4}, 
which are coupled by the chromomagnetic term of the interacting potential.
Therefore, in this case one has to solve the coupled equations,
\begin{eqnarray}
t_{11} &=& V_{11}+V_{11}G_0t_{11}+V_{12}G_0t_{21} \, ,
\nonumber\\
t_{21} &=& V_{21}+V_{21}G_0t_{11}+V_{22}G_0t_{21} \, ,
\label{eq19}
\end{eqnarray}
where the diagonal interactions $V_{11}$ and $V_{22}$ show contributions
from the chromoelectric and chromomagnetic terms of the interaction, while
the off-diagonal interactions $V_{12}$ and $V_{21}$ contain only the
contribution of the chromomagnetic part of the interacting potential.
As expected, the confinement and Coulomb terms are the dominant ones such that the
spin-spin term is just a small perturbation. The effect of the
non-diagonal terms is very small and it can be safely neglected.

\section{Results}
\label{secIV}
\begin{figure}[t]
\vspace*{-.7cm}
\hspace*{-0.2cm}\includegraphics[width=.5\columnwidth]{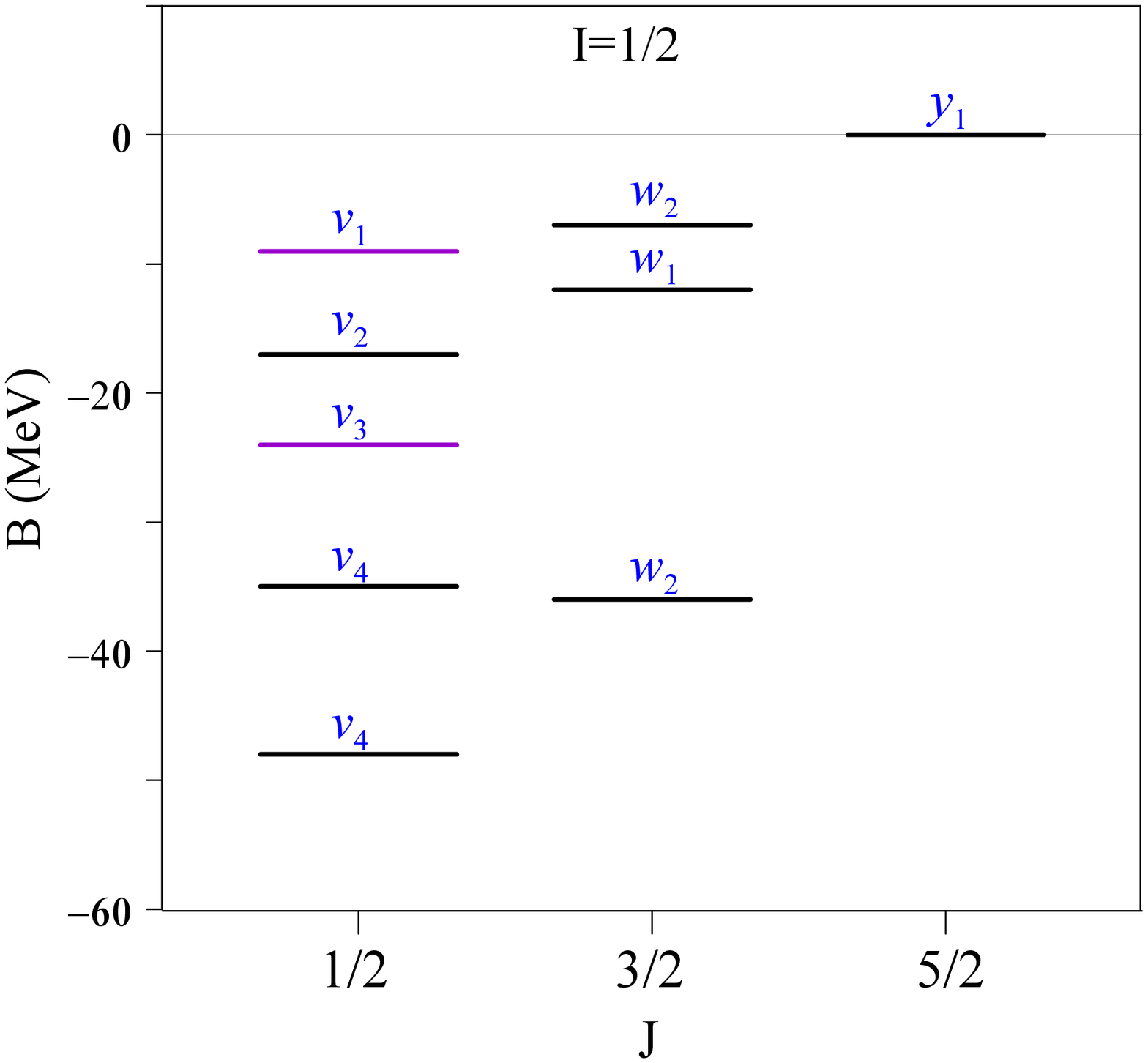}
\includegraphics[width=.5\columnwidth]{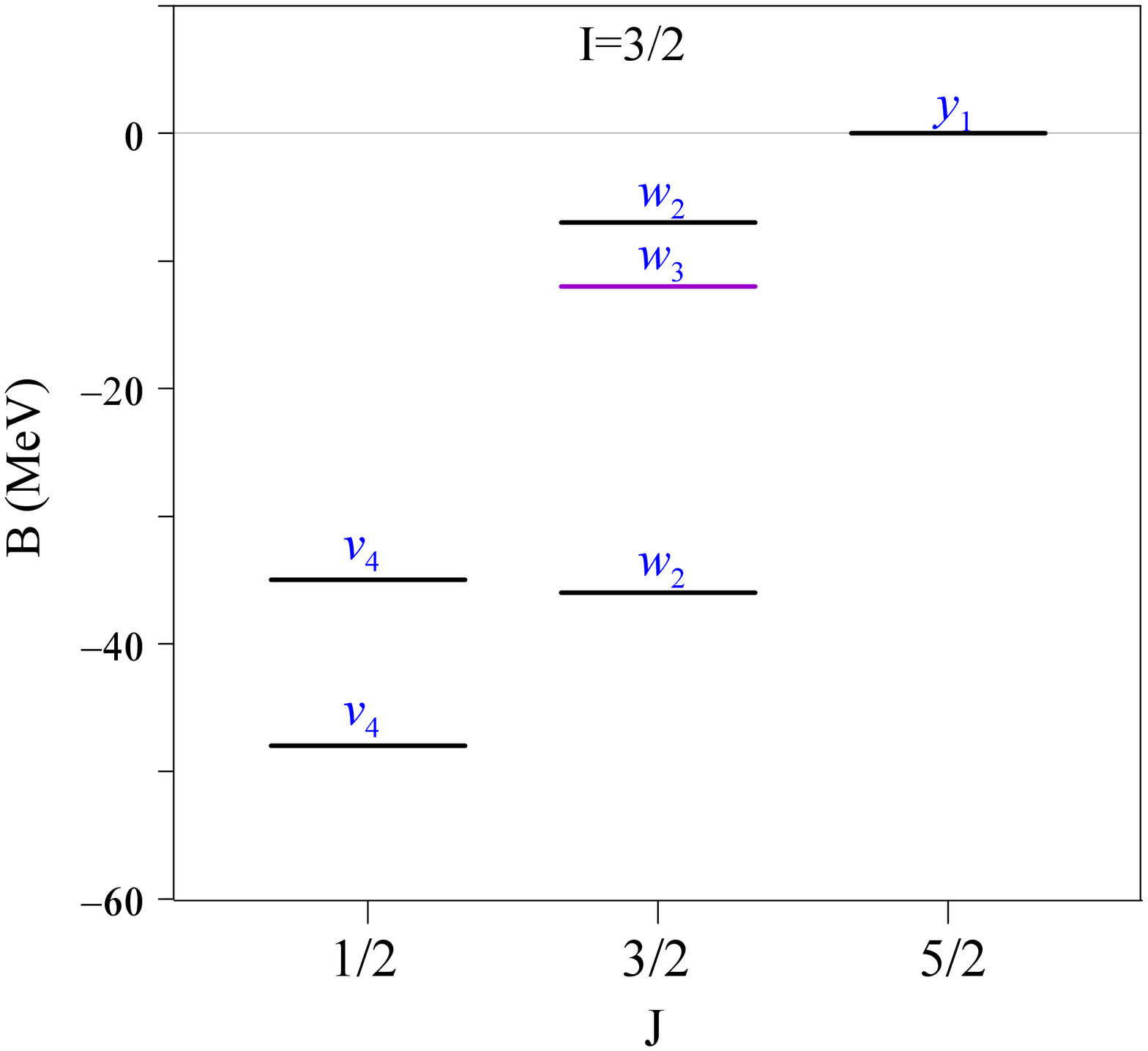}
\vspace*{-5cm}
\caption{Binding energy, in MeV, for the different hidden-charm 
pentaquarks. Black lines stand for states containing a spin one 
heavy quark-antiquark pair and purple lines denote states with a 
spin zero heavy quark-antiquark pair. The corresponding vector 
of Table~\ref{tab1} is indicated in the figure. 
Left panel: $I=1/2$. Right panel: $I=3/2$.}
\label{fig2}
\end{figure}

We have solved the three-body problem for the different $(I,J)$ 
states as discussed in Sec.~\ref{secIII} by taking
$m_1=m_{u,d}$, $m_2=2 m_c$ and $m_3=2 m_{u,d}$.~\footnote{It has been 
explicitly checked that the binding energy remains almost 
constant, it varies less than 1.5 MeV, for small variations, up to 10\%, of $m_2$ and $m_3$
around its central value.}
We show in Fig.~\ref{fig2} the five-quark states that are below threshold. Regarding 
the isospin 1/2 states, left panel, they are organized in two different 
shells. The lowest band contains $J=1/2$ and $3/2$ states with the 
two-quark subclusters with maximum spin. It is worth to note that 
in the two-baryon system the one-gluon exchange also generates the larger attraction 
for parallel spin configurations~\cite{Oka80,Gol89,Pan01,Mot99,Val01}.
Some states are rather close in energy and therefore hard to distinguish experimentally. 
In the upper shell there appear states with $J=1/2$, $3/2$ and $5/2$. 
The $J=5/2$ state is at threshold. Right panel 
on Fig.~\ref{fig2} shows the spectra of the isospin $3/2$ states.

Let us first note the degeneracy existing between $I=1/2$ and $I=3/2$ states, as could have been
expected a priori due to the isospin independence of the potential model in Eq.~(\ref{ecu3}),
although the result is not trivial due to the requirements of the Pauli principle. Secondly,
it has been checked that the conclusions dealing with stability or instability of multiquarks survive
variations of the parameters, we have specifically checked that the pattern remains for different
strengths of the spin-spin interaction by modifying the regularization parameter, $r_0$
in Eq.~(\ref{ecu3}).

\begin{figure}[t]
\vspace*{-.7cm}
\includegraphics[width=.55\columnwidth]{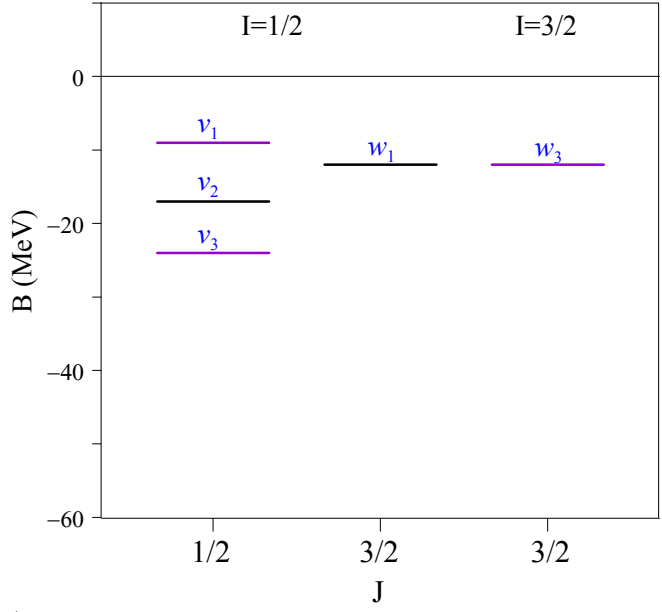}
\vspace*{-5.cm}
\caption{$I=1/2$ and $3/2$ hidden-charm pentaquarks containing substructures 
dictated by the quark correlations dominating the QCD phenomena~\cite{Jaf05}.
The notation is the same as in Fig.~\ref{fig2}.}
\label{fig4}
\end{figure}

There are additional quark correlations dominating the QCD phenomena~\cite{Jaf05}
that could hint to the most favorable states that can be observed in nature. First,
the very strong quark-antiquark correlation in the color-, flavor-, and spin-singlet 
channel $\{{\bf 1}_c, {\bf 1}_f, 0_s\}$ which can be viewed as 
the responsible for chiral symmetry breaking. The attractive forces in this channel are so
strong that condenses in the vacuum, breaking $SU(N_f)_L \times SU(N_f)_R$ chiral symmetry.
The next most attractive channel in QCD seems to be the color antitriplet, 
flavor antisymmetric, spin singlet $\{{\bf \bar 3}_c, {\bf \bar 3}_f, 0_s\}$, that would select
the $qq$ configurations most important spectroscopically. Thus, we show in Fig.~\ref{fig4} the resulting
spectrum by selecting those states that contain at least one the most attractive QCD channels,
i.e., a diquark with spin zero. It is observed that
the $J=5/2$ state at threshold disappears as well as the pentaquarks of the lowest shell. 

The general properties of the multiquarks favored by the quark correlations dominating 
the QCD phenomena shown in Fig.~\ref{fig4} can be easily estimated. 
In the charmonium sector, the mass difference between the $Q\bar Q$ $\{{\bf 1}_c, {\bf 1}_f, 1_s\}$
and $\{{\bf 1}_c, {\bf 1}_f, 0_s\}$ correlated states could be assimilated to the
$J/\Psi - \eta_c$ mass difference. The mass difference between the $qq$ $\{{\bf \bar 3}_c, {\bf 6}_f, 1_s\}$
and $\{{\bf \bar 3}_c, {\bf \bar 3}_f, 0_s\}$ has been estimated from full lattice QCD simulations
to be in the range of 100$-$200 MeV~\cite{Fra21,Ale06,Gre10}. We have tuned the effective masses of the correlated
structures to the hidden-charm pentaquarks, considering the following 
realistic values,
\begin{align}\label{mass}
\Delta M^{Q\bar Q} = M^{Q\bar Q}_{\{{\bf 1}_c, {\bf 1}_f, 1_s\}} - M^{Q\bar Q}_{\{{\bf 1}_c, {\bf 1}_f, 0_s\}} &= 86 \,\, {\rm MeV} \nonumber \, , \\
\Delta M^{qq} = M^{qq}_{\{{\bf \bar 3}_c, {\bf 6}_f, 1_s\}}  - M^{qq}_{\{{\bf \bar 3}_c, {\bf \bar 3}_f, 0_s\}} &= 146 \,\, {\rm MeV} \, .
\end{align}
Thus, denoting by $M_0$ the sum of the masses of a spin zero $Q\bar Q$ diquark, a spin zero $qq$ diquark
and a light quark, the mass of the states shown in Fig.~\ref{fig4} would be given by
\begin{equation}
M_i = M_0 - B_i + \Delta M^{Q\bar Q} \, \delta_{s_2,1} + \Delta M^{qq} \, \delta_{s_3,1} \, ,
\end{equation}
where $B_i$ is the binding energy calculated above. By taking $M_0=4321$ MeV,
one gets the results shown in Table~\ref{tab3}.
\begin{table}[t]
\begin{tabular}{cp{0.3cm}cp{0.3cm}cp{0.3cm}cp{0.3cm}c} 
\hline\hline
Vector  && $(I)J^P$     && $M_{\rm Th}$ (MeV) &&      State          && $M_{\rm Exp}$ (MeV)                \\ \hline
$v_1$   && $(1/2)1/2^-$ && 4312               &&       $P_c(4312)^+$ && $4311.9 \pm 0.7 ^{+6.8}_{-0.6}$     \\ 
$v_2$   && $(1/2)1/2^-$ && 4390               && \multirow{2}{*}{$P_c(4380)^+$} && \multirow{2}{*}{$4380 \pm 8 \pm 29$}             \\
$w_1$   && $(1/2)3/2^-$ && 4395               &&                                &&                       \\
$v_3$   && $(1/2)1/2^-$ && 4443               &&       $P_c(4440)^+$ && $4440.3 \pm 1.3 ^{+4.1}_{-4.7}$     \\ 
$w_3$   && $(3/2)3/2^-$ && 4455               &&       $P_c(4457)^+$ && $4457.3 \pm 0.6 ^{+4.1}_{-1.7}$    \\  
\hline\hline
\end{tabular}
\caption{Properties of the hidden-charm pentaquarks of Fig.~\ref{fig4}.}
\label{tab3}
\end{table}
As can be seen, there is a good agreement between theoretical states 
showing the most important correlations dictated
by the QCD phenomena and the experimental data~\cite{Aai15,Aai19}.
Thus, Table~\ref{tab3} presents a theoretical spin-parity assignment
for the existing hidden-charm pentaquarks. 

The spin-parities of the hidden-charm pentaquarks are not yet determined~\cite{Bra20}.
Nevertheless, there are predictions based on different models that can be compared with.
Although there are different proposals about the $P_c(4312)^+$ quantum numbers~\cite{Zhu16},
there seems to be a general preference for $J^P=1/2^-$~\cite{Wun10,Wan11,Yan12,Wul12,Xia13,Yaa17}, 
as it is found in our model. The two
narrow overlapping structures, $P_c(4440)^+$ and $P_c(4457)^+$~\cite{Aai19},
were originally reported as a single state, $P_c(4450)^+$~\cite{Aai15}. 
There were earlier predictions of two almost degenerate states with $J^P=1/2^-$ 
and $3/2^-$ at the position of the $P_c(4450)^+$ pentaquark. These structures
corresponded to the $J^P=1/2^-$ and $3/2^-$
hidden-charm states created dynamically by the $\Sigma_c \bar D^*$ charmed 
meson-baryon interactions~\cite{Wun10,Wum11,Xia13}. They were 
also predicted as bound states of charmonium $\Psi(2S)$ and the nucleon~\cite{Eid16}.
In both cases the quantum numbers of
the $P_c(4440)^+$ and $P_c(4457)^+$ pentaquarks
agree with our findings.
Finally, in our model there are two theoretical candidates, one with $J=1/2$ and the 
other with $J=3/2$, for the $P_c(4380)^+$, a wide resonance whose nature is still an intriguing issue
and is an outstanding challenge for future experiments~\cite{Kar15}.
The preferred spin assignment for this state was $J=3/2$ or $5/2$~\cite{Aai15}. A recent analysis
of $B_s \to J/\Psi p \bar p$ decays supports a $J^P=3/2^-$ assignment~\cite{Wag21}.
Thus, we could assign the $P_c(4380)^+$ to the $J^P=3/2^-$ state of 
our model and therefore leaving open the existence of another $J^P=1/2^-$
pentaquark in the same energy region, with a mass of about 4390 MeV.

Preliminary analysis of the experimental data suggested the coexistence 
of negative and positive parity pentaquarks in the same energy region~\cite{Aai15}.
We have studied such possibility within our model.
For this purpose, we have calculated the mass of the lowest positive parity
state, the first orbital angular momentum excitation of the $v_1$ state.
The technical details have been described in Sec.~\ref{subsecIII.2}. 
We chose this state because it is made up of the most strongly correlated structures, 
$Q\bar Q$ $\{{\bf 1}_c, {\bf 1}_f, 0_s\}$ and
$qq$ $\{{\bf \bar 3}_c, {\bf \bar 3}_f, 0_s\}$.
Then, it might have a similar mass to negative parity states 
made up of spin 1 structures. We have obtained an
energy of 197 MeV above threshold. By using the values given in Eq.~(\ref{mass})
one obtains a mass of 4518 MeV for two degenerate states with quantum numbers
$J^P=1/2^+$ and $3/2^+$. Therefore, positive parity pentaquark states
would appear above 4.5 GeV, a mass slightly larger than that of the states measured so far.
Similarly, most of the theoretical works prefer to 
assign the lowest lying pentaquarks to negative parity states.
Almost degenerate negative and positive parity states
may occur for hidden-flavor pentaquarks that have been detected 
in the same channel but that were formed by different pairs 
of quarkonium-nucleon states~\cite{Eid16}, one of them radially excited. Thus the negative 
parity pentaquark of the $(Q\bar Q)_{n+1, S}(qqq)$\footnote{$n$ stands for the 
radial quantum number of the $Q\bar Q$ system.} system would 
have a similar mass than the positive parity orbital angular 
momentum excited state of the $(Q\bar Q)_{n, S}(qqq)$ system. The assignment
of negative and positive parity states to different parity 
Born-Oppenheimer multiplets has already been suggested as a 
plausible solution in the triquark-diquark 
picture of Ref.~\cite{Gir19}. Nevertheless, this issue remains 
one of the most challenging problems in the pentaquark phenomenology
that should be first confirmed experimentally.
 
Multiquark states would show very different decay patterns regarding 
its internal structure~\cite{Wul17}. The decays of the pentaquarks in
Table~\ref{tab3} into an (anti)charmed meson + charmed baryon are 
strongly suppressed since decays into open charm channels can go 
only via $t$-channel exchange by a heavy $D$ meson. Due to the content 
of the pentaquarks states they would follow the decays of charmonium 
excited states, $\Psi(nS)$ and $\eta_c(nS)$. Thus, multiquarks containing 
a spin zero heavy quark-antiquark pair: $v_1$, $v_3$ and $w_3$
in Table~\ref{tab3}, would be narrower than those with a spin one 
heavy quark-antiquark pair: $v_2$ and $w_1$ in Table~\ref{tab3}. 
This corresponds nicely with the experimental 
observations. However, besides the contribution to the width 
of the substructures that form each pentaquark, one should also 
consider the width due to the bound nature of the system. At this 
point it is worth to mention that the final width of a resonance 
does not come only determined by its internal content, but there 
are significant corrections due to an interplay between the phase 
space for its decay to the detection channel and its mass with 
respect to the hadrons generating the state~\cite{Gar18}.

The existence of hidden-flavor pentaquarks has been concluded in various 
constituent quark-model studies. Let us analyze our results compared to 
other related approaches in the literature. Ref.~\cite{Hua16} 
studies hidden-charm pentaquarks in a quark delocalization color screening model, where besides 
the one-gluon exchange potential quarks interact through the exchange of Goldstone bosons.
It presents results for $I=1/2$ pentaquarks concluding the existence of several 
negative parity bound states with
$J=1/2$, $3/2$, and $5/2$. The lowest state corresponds to $J=1/2$ and the $J=5/2$ state
is at the edge of binding. The deepest states with $J=1/2$ and J=$3/2$ are found in the
$(Q\bar Q)(qqq)$ configuration. This is the structure favored by the color
Coulomb-like short-range correlations between the heavy flavors. In fact, the
$(q\bar Q)(Qqq)$ configuration only shows quasi-stable states that should be confirmed 
by investigating the scattering process of other open channels. The results are in good 
agreement with the $I=1/2$ results of our model, the deepest 
state being $J=1/2$ while the only $J=5/2$ state is at threshold. These findings come to 
give support to the existence of a color-singlet correlation between the heavy flavors 
within the pentaquarks. 

Ref.~\cite{Yan17} discusses results for $I=1/2$ states based on a chiral-quark model. 
Several negative parity bound states with $J=1/2$, $3/2$, and $5/2$ are reported.
In contrast to Ref.~\cite{Hua16} the dominant configuration is found to be $(q\bar Q)(Qqq)$.
It may be because in the $(Q\bar Q)(qqq)$ configuration, quarkonium and 
baryons do not share light $u$ and $d$ quarks and thus the OZI rule suppresses the 
interactions mediated by the exchange of mesons made of only light 
quarks~\cite{Bro90}.\footnote{This could also be the reason why the hadronic molecular scenario
prefers to describe hidden-flavor pentaquarks as bound states of open-flavor
hadrons. See Refs.~\cite{Hos16,Guo18,Xia19,She19,Wan20,Bur15}.} The exchange of $D$ mesons is 
a too short-range interaction to compete with the medium-range attraction that
can be generated by light-meson exchanges arising in the $(q\bar Q)(Qqq)$ configuration.
It is also worth to note that hybrid models containing gluon and meson exchanges
at quark level show a reduced strength of the one-gluon exchange potential~\cite{Vij04}. This is
because pseudoscalar meson exchanges between quarks do also contribute to 
the $\Delta - N$ mass difference. However, the pseudoscalar spin-flavor interaction 
favors different color-spin components than those favored by
the one-gluon exchange~\cite{Val97}. As a consequence, a distinct
pattern of multiquark states is found in hybrid or pure one-gluon exchange
approaches.  

Different studies concluded the existence of $I=3/2$ pentaquarks.
In the so-called hadroquarkonium approach, Ref.~\cite{Per16} presents 
robust predictions of isospin $3/2$ bound states 
of $\Psi(2S)$ and $\Delta$ with masses around 4.5 GeV.
Looking back to constituent quark approaches, Ref.~\cite{Ric17} concluded 
the existence of $I=3/2$ hidden-flavor pentaquarks. The
pattern obtained is rather similar to our
calculation, with the $J=5/2$ state being almost at threshold (note that
we only consider relative $S$ waves). Regarding the $I=1/2$ states, it
is the presence of the $(q\bar Q)(Qqq)$ configuration, in other words 
repulsive color octets in the $(Q\bar Q)(qqq)$ configuration, which rules out
the possibility of having bound states. Therefore, the dynamical 
correlations arising among the heavy flavors are more effective for 
isospin $1/2$ states. This is easily understandable due to the fully
symmetric nature of the isospin $3/2$ wave function, which in itself
reduces the allowed Hilbert space vectors.

Ref.~\cite{Wul17} studies hidden-flavor pentaquarks using the more
repulsive color octet-color octet component, ${\bf 8}_{(Q\bar Q)} {\bf 8}_{(qqq)}$.
A set of negative parity states that would remain bound only against the 
heavier $(q\bar Q)(Qqq)$ threshold is reported. The most distinctive feature
of this approach lies in the fact that compact pentaquarks with a colored 
$qqq$ cluster have small branching ratios for the hidden-flavor decay channels
as compared to possible baryon-meson molecules.

Ref.~\cite{Wen19} makes use of an extended chromomagnetic model where besides
the color-spin chromomagnetic potential, effective quark-pair mass parameters
accounting for the effective quark masses and the color interaction between two 
quarks are considered. These parameters are fitted to the meson and baryon
spectra. 10 $I=1/2$ and 7 $I=3/2$ hidden-flavor pentaquarks are found.
All of them are negative parity states and
there appear $J=1/2$, $3/2$ and $5/2$ pentaquarks in both isospin channels. 
The pattern of $I=1/2$ states shown in Fig. 1 of Ref.~\cite{Wen19} is 
similar to our results. However, the degeneracy between
$I=1/2$ and $I=3/2$ states is not observed in the spectra. This could be
due to the way the effective quark-pair mass parameters are determined,
because the interacting potential is isospin independent.  

In addition to the models we have discussed with which the comparison is meaningful
since they follow a similar constituent approach, as mentioned in the introduction,
there are different proposals used to study hidden-flavor pentaquarks. The predictions 
of diquark models are very varied~\cite{Mai15,Gir19,Ali19,Shi21}, depending on the 
hypotheses used for the diquark dynamics. Some further assumptions are
sometimes made about the chromomagnetic interaction between diquarks~\cite{Mai14}. 
A similar general conclusion can be derived from QCD sum rules studies, where
one can find either molecular approaches~\cite{Zha19,Che19,Azi17,Wan21}
or others based on hidden-color components~\cite{Pim20}.
A recent review about the status of heavy quark sum rules
and the uses for exotic hadron molecules can be found in Ref.~\cite{Nar21}.
Hadronic molecular models based either on effective chiral Lagrangians 
or one-boson exchange potentials rely on the determination of unknown 
low-energy parameters and coupling constants, the latter usually determined by 
quark-model relations~\cite{Wun10,Yam17,Men19,Wan19,Yal21}.
Predictions obtained under the hypothesis about the structure of some
of the novel states, used to fix the unknown constants, are a nice tool
to analyze forthcoming states in the heavy-hadron spectra.
Generally speaking it can be said that in all approaches the resulting 
spectra are very rich. For a more detailed analysis 
of the particularities of each approach we refer the reader to the aforementioned
reviews~\cite{Che16,Bri16,Ric16,Hos16,Che17,Leb17,Ali17,Esp17,Guo18,Ols18,Kar18,Bra20}
and references therein.

The model we explore has a well defined asymptotic threshold made of a light baryon, $N$ or
$\Delta$ depending on the isospin, and a vector or a pseudoscalar quarkonium state, 
depending on the spin component of the heavy quark-antiquark pair. In contrast to 
molecular hadronic models based on effective interactions
between hadrons, the approach we follow could be generalized to any other hidden-flavor 
system without the need of additional ingredients. Our study is just based on
the correlations dictated by the QCD dynamics on a realistic quark-quark 
interaction, see Eq.~(\ref{ecu3}), that describes the low-energy baryon and
meson spectra, see Table~\ref{tab2}. It is worth noting that the correlations 
used do not lead to stable multiquarks for any quark substructure, in the same way
the $NN$ short-range repulsion induced by the one-gluon exchange dynamics is not
universal and disappears for other two-hadron channels. Thus, for example, the
QCD correlations used in this work would not constraint the color wave function of
pentaquarks with anticharm or beauty, $\bar Q qqqq$. Therefore, such systems would not
present bound states, as recently discussed in Ref.~\cite{Ric19}, due to a non favorable 
interplay between chromoelectric and chromomagnetic effects.

Finally, the results we have presented could be further used to study the 
possible existence of charmonium states bound to atomic nuclei suggested 
by Brodsky~\cite{Bro90} more than three decades ago. As it has been 
mentioned above, since charmonium and nucleons do not share light $u$ and $d$ quarks, 
the OZI rule suppresses the interactions mediated by the exchange of mesons made of only light 
quarks. Thus, if such states are indeed bound to nuclei, it has been emphasized the relevance
to search for other sources of attraction~\cite{Kre18}. 
A charmonium-nucleon interaction which provides a binding mechanism has been found,
in the heavy-quark limit, in terms of charmonium chromoelectric 
polarizabilities and densities of the nucleon energy-momentum tensor~\cite{Eid16,Per16,Dub08}.
The existence of such bound states has also 
been justified by changes of the internal structure of the hadrons in the nuclear medium.
Thus, for example, $J/\Psi$-nuclei bound states were found in Ref.~\cite{Tsu11}.
In a similar model it has been recently concluded that the $\eta_c$ meson should form bound states 
with all the nuclei considered, from $^4$He to $^{208}$Pb~\cite{Cob20}.
Our model presents an alternative mechanism based on the 
short-range one-gluon exchange interaction between the constituents of charmonium and 
nucleons. This mechanism has already been suggested to lead to dibaryon 
resonances~\cite{Oka80,Gol89,Pan01,Mot99,Val01,Cle17}.
To our knowledge, this result has never been obtained before based on 
pure quark-gluon dynamics using a restricted Hilbert space.

\section{Summary}
\label{secV}

In short, we have studied hidden-flavor pentaquarks imposing the dynamical
correlations inherent to the color Coulomb-like nature of the short-range one-gluon exchange
interaction. Such correlations lead to a frozen color wave function of the five-body
system, which allows to reduce the problem to a more tractable three-body problem.
The three-body problem has been exactly solved by means of the Faddeev equations.
To perform exploratory studies of systems with more than 
three-quarks it is of basic importance to work with models that correctly
describe the two- and three-quark problems of which thresholds are made 
of. Thus, the interactions between the constituents are deduced from a generic constituent 
model, the AL1 model, that gives a nice description of the low-energy baryon and meson
spectra. 

The dynamical correlations arising from the one-gluon exchange interaction
due to the presence of a heavy quark-antiquark pair result in several
bound states. The lightest pentaquarks have $J=1/2$ 
and $3/2$. $J=5/2$ states lie at threshold. Under the assumption 
that nature favors multiquarks which are made up of correlated 
substructures dictated by QCD, we have estimated the mass of 
the lowest lying pentaquarks. We have considered realistic values 
for the mass difference of the correlated quark pairs. A good
description of the experimental data has been obtained.
The tentative spin-parity assignment of the different pentaquarks
agrees well with other approaches dedicated to study a particular 
set of states. 

Our study is just based on the correlations dictated by the QCD 
dynamics on a realistic quark-quark interaction. Thus, it could 
be generalized to any other hidden-flavor system without the need 
of additional ingredients. It is worth noting that the correlations 
used do not lead to stable multiquarks for any quark substructure.
Thus, for example, the QCD correlations used in this work would 
not constraint the color wave function of pentaquarks with anticharm 
or beauty.

As a bonus of our calculation we have found a dynamical model that would
account for the existence of quarkonium states bound to nuclei.
The existence of such bound states has been justified in the hadrocharmonium
approach or by changes of the 
internal structure of the hadrons in the nuclear medium but, to our knowledge, 
it has never been obtained before based on pure quark-gluon dynamics 
using a restricted Hilbert space.

Bound states and resonances are usually very sensitive to model 
details and therefore theoretical investigations with different 
phenomenological models are highly desirable. We have tried to minimize the
influence of the interacting potential by using a standard constituent model and
we have explored the consequences of dynamical correlations arising from the
Coulomb-like nature of the short-range potential. Similar arguments were used in the
past to select dibaryon channels that might lodge resonances with success.
The pattern obtained could be scrutinized against the future experimental results
providing a great opportunity for extending our knowledge to some unreached 
part of the hadron spectra. More such exotic baryons are expected and needed to make
reliable hypotheses on the way the interactions in the system are
shaping the spectra.

\section{Acknowledgments}
A.V. thanks Prof. G.~Krein for valuable discussions about
the possibility of quarkonium states bound to nuclei. 
This work has been partially funded by COFAA-IPN (M\'exico) and 
by Ministerio de Ciencia e Innovaci\'on and EU FEDER under 
Contracts No. PID2019-105439GB-C22 and RED2018-102572-T.
\appendix
\section{Coupling of different Faddeev amplitudes}
\label{AppA}
In the Faddeev formalism the amplitude $T_\alpha$ is coupled to the amplitudes
$T_\beta$, with $\beta \ne \alpha$, corresponding to different coupling schemes. In the
case of the spin part the wave functions are,
\begin{align}
|\alpha> &=|[(s_j,s_k)S_{jk};s_i]S>=\sum_{\mu_i\mu_j\mu_k}
C^{s_js_kS_{jk}}_{\mu_j,\mu_k}C^{S_{jk}s_iS}_{\mu_j+\mu_k,\mu_i}
|s_i\mu_i>|s_j\mu_j>|s_k\mu_k> \nonumber \, , \\
|\beta> &=|[(s_k^\prime,s_i)S_{ki};s_j]S>=\sum_{\nu_i\nu_j\nu_k}
C^{s_k^\prime s_iS_{ki}}_{\nu_k,\nu_i}C^{S_{ki}s_jS}_{\nu_k+\nu_i,\nu_j}
|s_i\nu_i>|s_j\nu_j>|s_k^\prime\nu_k> \, ,
\label{eq108}
\end{align}
so that the recoupling coefficients are
\begin{equation}
<\alpha | \beta >=\sum_{\substack{\mu_i\mu_j\mu_k \\
\nu_i\nu_j\nu_k}}
C^{s_js_kS_{jk}}_{\mu_j,\mu_k}C^{S_{jk}s_iS}_{\mu_j+\mu_k,\mu_i}
C^{s_k^\prime s_iS_{ki}}_{\nu_k,\nu_i}C^{S_{ki}s_jS}_{\nu_k+\nu_i,\nu_j}
<s_i\mu_i|s_i\nu_i><s_j\mu_j|s_j\nu_j><s_k^\prime\mu_k|s_k\nu_k> \, ,
\label{eq109}
\end{equation}
where,
\begin{align}
<s_i\mu_i|s_i\nu_i> &=\delta_{\mu_i\nu_i} \nonumber \, , \\
<s_j\mu_j|s_j\nu_j> &=\delta_{\mu_j\nu_j} \nonumber \, , \\
<s_k^\prime\mu_k|s_k\nu_k> &=\delta_{s_k^\prime s_k}\delta_{\mu_k\nu_k} \, ,
\label{eq112}
\end{align}
so that 
\begin{equation}
<\alpha |\beta >=\delta_{s_k s_k^\prime}\sum_{\mu_i\mu_j\mu_k}
C^{s_js_kS_{jk}}_{\mu_j,\mu_k}C^{S_{jk}s_iS}_{\mu_j+\mu_k,\mu_i}
C^{s_ks_iS_{ki}}_{\mu_k,\mu_i}C^{S_{ki}s_jS}_{\mu_k+\mu_i,\mu_j} \, .
\label{eq113}
\end{equation}
Thus, the recoupling coefficient $<\alpha |\beta >=0$ if $s_k\ne s_k^\prime$,
which leads to the decoupling of amplitudes when the spin of one particle is 
different in the states $|\alpha >$ and $| \beta >$.

\end{document}